\documentclass{svproc}

\usepackage{url}

\usepackage{graphicx}
\usepackage{epstopdf,times,color}
\usepackage{bm}
\usepackage{footnpag}

\begin{document}
\mainmatter
\title{Uncertainty Propagation Using Hybrid Methods}
\titlerunning{Uncertainty Propagation Using Hybrid Methods}
\author{Juan F\'elix San-Juan\inst{1} \and Montserrat San-Mart\'in\inst{2} \and Iv\'an P\'erez\inst{1} \and Rosario L\'opez\inst{1} \and Edna Segura\inst{3} \and Hans Carrillo\inst{3}}
\authorrunning{Juan F\'elix San-Juan et al.}
\institute{Scientific Computing Group (GRUCACI), University of La Rioja, \\26006 Logro\~no, Spain\\
\email{\{juanfelix.sanjuan,ivan.perez,rosario.lopez\}@unirioja.es}
\and
Scientific Computing Group (GRUCACI), University of Granada,\\  52005 Melilla, Spain\\
\email{momartin@ugr.es} 
\and
Department of Mathematics and Computer Science, University of La Rioja, \\26006 Logro\~no, Spain\\
\email{\{edna-viviana.segura,hans-mauricio.carrillo\}@alum.unirioja.es}
}

\maketitle

\noindent
This is an Author Accepted Manuscript version of the following chapter: Juan F\'elix San-Juan, Montserrat San-Mart\'in, Iv\'an P\'erez, Rosario L\'opez, Edna Segura, and Hans Carrillo, ``Uncertainty Propagation Using Hybrid Methods,'' published in International Conference on Soft Computing Models in Industrial and Environmental Applications (SOCO 2020), edited by \'Alvaro Herrero, Carlos Cambra, Daniel Urda, Javier Sedano, H\'ector Quinti\'an, and Emilio Corchado, 2021, Springer, reproduced with permission of Springer Nature Switzerland AG. The final authenticated version is available online at: \url{http://dx.doi.org/10.1007/978-3-030-57802-2_68}.

Users may only view, print, copy, download and text- and data-mine the content, for the purposes of academic research. The content may not be (re-)published verbatim in whole or in part or used for commercial purposes. Users must ensure that the author's moral rights as well as any third parties' rights to the content or parts of the content are not compromised.

\begin{abstract}
Small corrections in the argument of the latitude can be used to improve the accuracy of the SGP4 orbit propagator. These corrections have been obtained by applying the hybrid methodology for orbit propagation to SGP4, therefore yielding a hybrid version of this propagator. The forecasting part of the hybrid method is based on a state-space formulation of the exponential smoothing method. If the error terms that have to be considered during the model fitting process are taken as Gaussian noise, then the maximum-likelihood method can be applied so as to estimate the parameters of the exponential-smoothing model, as well as to compute the forecast together with its confidence interval. Finally, this hybrid SGP4 orbit propagator has been applied to data from Galileo-type orbits. This new propagator improves the accuracy of the classical SGP4, especially for short forecasting horizons.
\keywords{Time series, Hybrid methodology, Orbit propagation, Uncertainty.}
\end{abstract}

\section{Introduction}

The orbital motion of an artificial satellite, or space-debris object, is influenced by a variety of external perturbations, in addition to the Earth's gravity, which is the principal force that determines its orbit, such as the atmospheric drag, third-body influences, the solar radiation pressure, the Earth's tidal effects, and, in the case of an artificial satellite, also the small perturbing forces produced by its propulsion system. Numerical, analytical or semi-analytical methods can be used in order to solve the nonlinear equations of motion of this complex dynamical system. With the aim of simplifying it, some of  the aforementioned external forces may be ignored depending on the intended purpose, for example the scientific requirements for the mission of an Earth's satellite, or the maintenance of a space-debris catalog. An orbit propagator is the implementation of one of the aforementioned solutions as a computer program.

The maintenance of a running catalog of space objects orbiting the Earth is an unavoidable duty in the management of the space environment close to the Earth, which requires the orbital propagation of tens of thousands of objects. Currently, these ephemerides are publicly available through the North American Aerospace Defense Command (NORAD) catalog, yet other organizations, like the European Space Agency (ESA), may make their own data, obtained from observations, also accessible.

Due to the huge number of objects to be propagated, a compromise between accuracy and efficiency must be established, depending on a variety of criteria. High-fidelity propagation models usually require step-by-step propagation by using numerical methods, which are computationally intensive because they rely on small step sizes. On the other hand, simplified models may admit analytical solutions, in this way notably alleviating the computational burden. In either case, the orbit propagation program relies only on the initial conditions, as well as on the propagation model, to make its predictions. However, the collection of past ephemerides provided by the catalog can be used to improve orbit predictions by taking non-modeled effects into account.

The main application of a space-debris catalog is the forecast of the future positions of all cataloged objects, since their extreme velocity converts them into uncontrolled projectiles that pose a real threat to operative satellites and space assets. As a result of this massive propagation activity, collision warnings have to be broadcast, so that satellite operators can perform collision-avoidance maneuvers. The assessment of the collision risk is strongly affected by all the uncertainties involved in the process of predicting the future positions of the cataloged objects.

The hybrid methodology for orbit propagation allows combining a classical propagation method, which can be numerical, analytical or semi-analytical, and a forecasting technique, based on either statistical time-series models \cite{san2012gru_sarimahop} or machine-learning techniques, which is able to generate a compensation for the classical-propagation future errors from the time series of its former errors. This combination leads to an increase in the accuracy of the base propagator for predicting the future position and velocity of an artificial satellite or space-debris object, since it allows modeling higher-order terms and other external forces not considered in the base propagator.

In this work, we make use of a hybrid approach which combines the well-known analytical orbit propagator \textit{Simplified General Perturbations-4} (SGP4), specially designed to be used with \textit{Two-Line Elements} (TLE) as initial conditions \cite{hoo1980_spacetrack3,val2006_spacetrack3rev,san2016gru_hsgp4_aas,san2016gru_hsgp4_icatt,san2017gru_hsgp4}, with a state-space formulation of the exponential smoothing method \cite{hyn2002_expsmstsp_art,sny2002_forecexpsmooth,hyn2005_expsmstsp_art,hyn2008_expsmstsp_book}. The consideration of the error terms as Gaussian noise during the model fitting process allows us to use the maximum likelihood method to estimate the parameters of the exponential smoothing model, as well as to compute the forecast and its confidence interval. Our goal in this study is to verify the capability of the hybrid orbit propagator to propagate the initial uncertainty.

The outline of the paper is structured around three sections. The hybrid methodology for orbit propagation is concisely summarized in Section 2. Then, the application of the hybrid SGP4 propagator to Galileo-type orbits is discussed in Section 3. Finally, Section 4 draws the conclusions of the study.

\section{Hybrid Methodology}

The hybrid methodology for orbit propagation is aimed at improving the estimation of the future position and velocity of any artificial satellite or space-debris object at a final instant $t_f$, expressed in some set of canonical or non-canonical variables, $\hat{\mathbf{x}}_f$. That improvement is performed on an initial approximation $\mathbf{x}_f^{\mathcal{I}}$, obtained by means of a base propagator that applies an integration method $\mathcal{I}$, which can be numerical, analytical or semi-analytical, to the system of differential equations that govern the behavior of the nonlinear dynamical system.

In order to enhance this initial approximation, it is necessary to somehow know the dynamics that the base propagator is missing. For that purpose, we can use the time series of its former errors, for which we need to know the real satellite ephemerides, either obtained by observation or simulated by high-fidelity slow numerical propagation, during a past \textit{control interval}. For every epoch $t_i$ in this control interval, we calculate the error $\mathcal{\bm\varepsilon}_i$ as the difference between the real ephemeris $\mathbf{x}_i$ and the base-propagator approximation $\mathbf{x}_i^{\mathcal{I}}$:

\begin{equation}
\mathcal{\bm\varepsilon}_i = \mathbf{x}_i - \mathbf{x}_i ^{\mathcal{I}}. \label{error} 
\end{equation}

This error $\mathcal{\bm\varepsilon}_i$ is, in part, due to the fact that the base propagator implements a simplified model of the real system, although the intrinsic error in the initial conditions that we want to propagate can also contribute to it.

Once we have the time series of the base-propagator former errors, which embeds the dynamics that we want to reproduce, we can apply statistical time-series methods or machine-learning techniques in order to build a model. Later, we will use that model to predict an estimation of the base-propagator error at the final instant $t_f$, $\hat{\mathcal{\bm\varepsilon}}_f$. Finally, the enhanced ephemeris at $t_f$, $\hat{\mathbf{x}}_f$, will be calculated by adding this estimated error to the base-propagator approximation $\mathbf{x}_f^{\mathcal{I}}$:

\begin{equation} \label{forecast}
\hat{\mathbf{x}}_f = \mathbf{x}_f^{\mathcal{I}} + \hat{\mathcal{\bm\varepsilon}}_f.
\end{equation}

\section{Application of the Hybrid SGP4 Propagator to Galileo-Type Orbits}

\subsection{SGP4 and AIDA Orbit Propagators}

Two orbit propagators are involved in this study: SGP4, which is the base propagator whose accuracy we intend to improve, and AIDA  \cite{mor2014gru_aida}, a high-precision numerical propagator that we use for generating the so-called \textit{pseudo-observations} that represent the satellite real ephemerides.

SGP4 is an analytical orbit propagator originally based on Brouwer's theory \cite{bro1959_astnodrag} of satellite motion perturbed by the first five zonal harmonics of the Earth gravitational field. The description of the original Fortran code can be found in \cite{hoo1980_spacetrack3}, although the complete documentation of all the mathematical equations was published in 2004 \cite{hoo2004_histanalUS}. In this work, we use the most updated code developed by Vallado \cite{val2006_spacetrack3rev}, which merges SGP4/SDP4 models, and is simply referred to as SGP4. This propagator includes the following force models: $J_2$ to $J_4$ zonal harmonics, air drag, and lunar and solar perturbations, as well as long-period resonant harmonics for the so-called deep-space satellites.

The input to the SGP4 propagator is the Two-Line-Element (TLE) set, which provides position and velocity vectors at a given time. The TLE includes information about the satellite and its orbit, such as the satellite number, orbit inclination, eccentricity, argument of perigee, derivatives of the mean motion, the BSTAR drag parameter, mean anomaly, and others.

The other orbit propagator, which we use for simulating observational data, is AIDA, the \textit{Accurate Integrator for Debris Analysis}. It includes the following force models:
\begin{itemize}
\item geopotential acceleration computed using the EGM2008 model \cite{pav2012_egm2008}, up to an arbitrary degree and order for the harmonics;
\item atmospheric drag, modeled using the NRLMSISE-00 air density model \cite{pic2002_nrlmsise-00};
\item solar radiation pressure with dual-cone shadow model;
\item third body perturbations from Sun and Moon.
\end{itemize}

\subsection{Numerical Results}

\begin{figure}
\centering
\includegraphics[scale=.5]{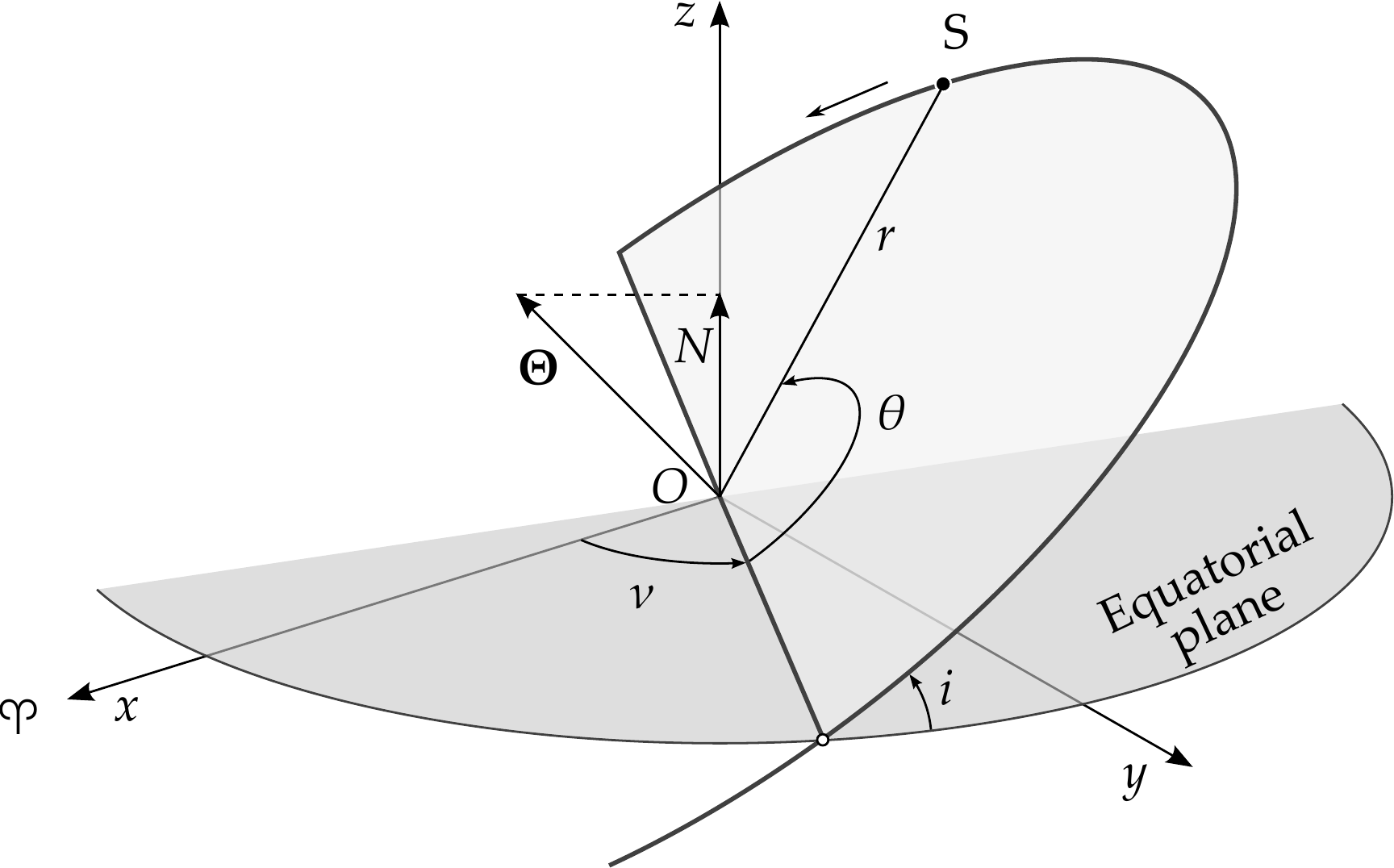}
\caption{Polar-nodal variables $(r,\theta,\nu,R,\Theta,N)$}
\label{fig1}
\end{figure}

This study has been conducted in the polar-nodal coordinates. The meaning of these variables is shown in Fig. \ref{fig1}. O$xyz$ represents an inertial reference frame centered at the center of mass of an Earth-like planet. The variable $r$ denotes the distance from the center of mass of the Earth-like planet to the space object S, $\theta$ is the argument of the latitude of the object, $\nu$ represents the right ascension of the ascending node, $R$ is the radial velocity, $\Theta$ designates the magnitude of the angular momentum vector $\bm{\Theta}$, whereas $N$ represents the projection of $\bm{\Theta}$ onto the $z$-axis.

In this study, the hybrid methodology has been applied only to the argument of the latitude $\theta$.

\begin{figure}[h!!]
\centering
\includegraphics[scale=.6]{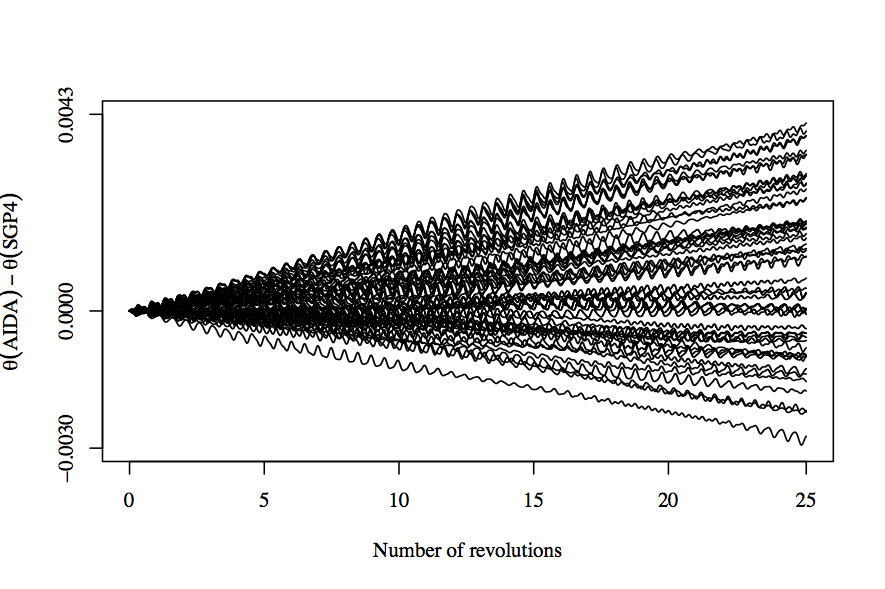}
\caption{$\varepsilon^\theta = \theta_{AIDA} - \theta_{SGP4}$ time series for several TLEs from  the  Galileo-8 satellite}\label{fig2}
\end{figure}

Fig. \ref{fig2} plots $\varepsilon^\theta=\theta_{AIDA}-\theta_{SGP4}$, the time series of the error in the argument of the latitude, for 53 different TLEs from the Galileo-8 satellite. TLE dates span from 28th March 2015 to 16th December 2016, including TLEs for every month between those two dates, with an approximately even distribution, although not completely regular. As can be seen, despite the fact that all these time series correspond to the same satellite, they do not seem to present a unique pattern. All of them show seasonal components, whose periods are approximately their Keplerian periods, and exhibit a high degree of variation in their trends.

\begin{figure}[h!!]
\centering
\includegraphics[scale=.6]{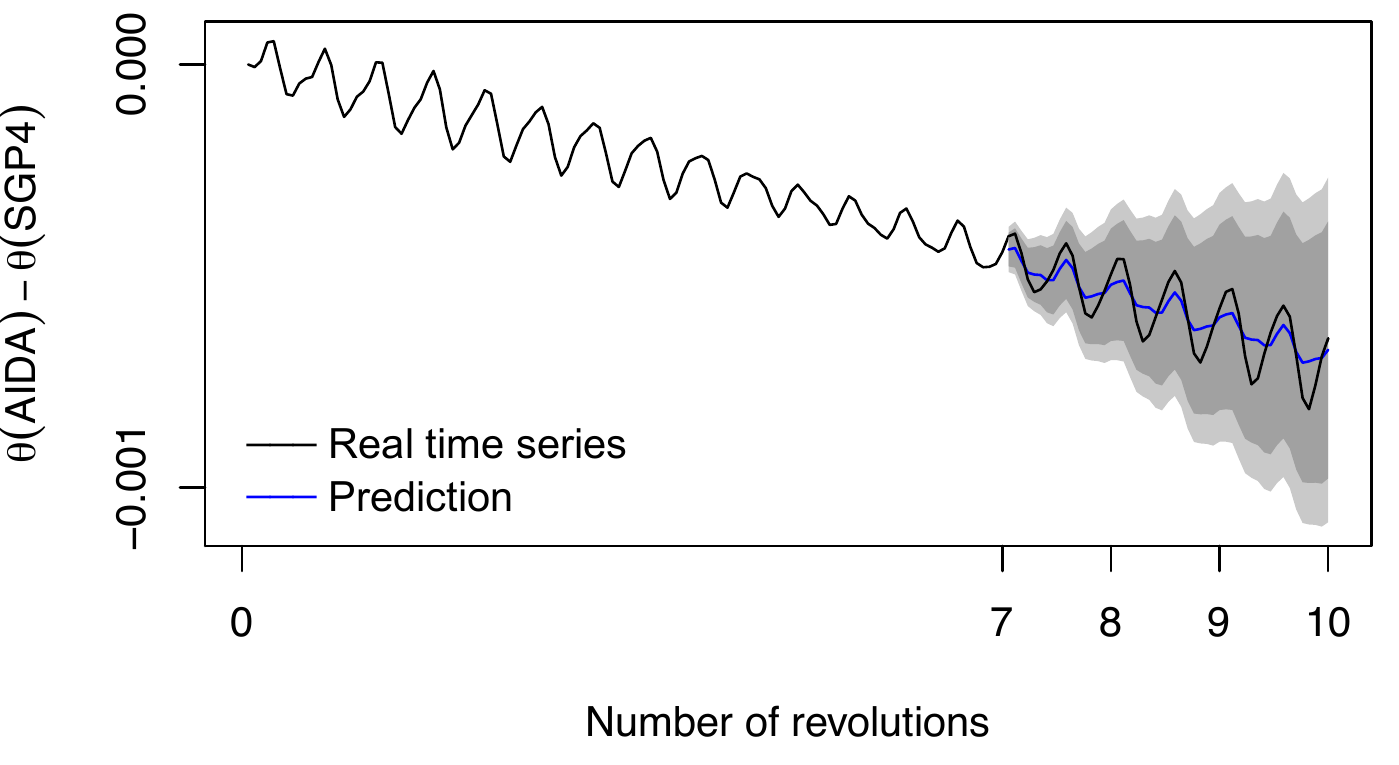}
\caption{Forecast of $\varepsilon^\theta$ for the next three satellite revolutions. The blue line represents the prediction and the shaded areas correspond to the 99\% and 95\% confidence intervals.}\label{fig3}
\end{figure}

Fig. \ref{fig3} displays the real and predicted values for one of the $\varepsilon^\theta$ time series shown in Fig. \ref{fig2} for the following TLE,\footnote{TLEs can be downloaded from  \url{https://www.space-track.org}.} which corresponds to 28th March 2015:
{\scriptsize
\begin{verbatim}
1 40545U 15017B   15087.10529976  .00000015  00000-0  00000+0 0  9997
2 40545 055.0895 094.8632 0005535 231.4671 034.4229 01.67457620    08
\end{verbatim}} 

Predictions have been generated by applying the state-space formulation of the exponential smoothing method. This formulation considers the error component as Gaussian white noise, which allows applying formal estimation techniques, such as the maximum-likelihood method. Under reasonable conditions, this method provides consistent estimations with asymptotic efficiency. In essence, the state-space formulation allows representing the evolution of a set of components that underlie the time series, although they are not directly observable. This method is based on the recursive application of two equations: the \textit{measurement equation}, which provides the estimation of the time-series values from the state vector, and the \textit{transition equation}, which defines the evolution of the state vector. A detailed description of this formulation can be found in Ref. \cite{hyn2002_expsmstsp_art}.

The first seven revolutions, which represent approximately $100$ hours, that is, about four days, constitute the control interval that we use for fitting the parameters of the model. Then, we use that model for predicting the time-series values during the next three revolutions. The line in black represents the real time series, whereas the line in blue corresponds to the forecast. As can be seen in the figure, the model works relatively well during the first forecasting revolutions. The use of the state-space formulation of the exponential smoothing method also allows determining the confidence interval of the prediction. The dark- and light-gray halos surrounding the forecast represent the 99\% and 95\% confidence intervals, respectively. It is worth noting that the size of the confidence interval is an indicator of the uncertainties associated to the fitting process of the statistical model.

\begin{figure}[h]
\centering
\includegraphics[scale=.6]{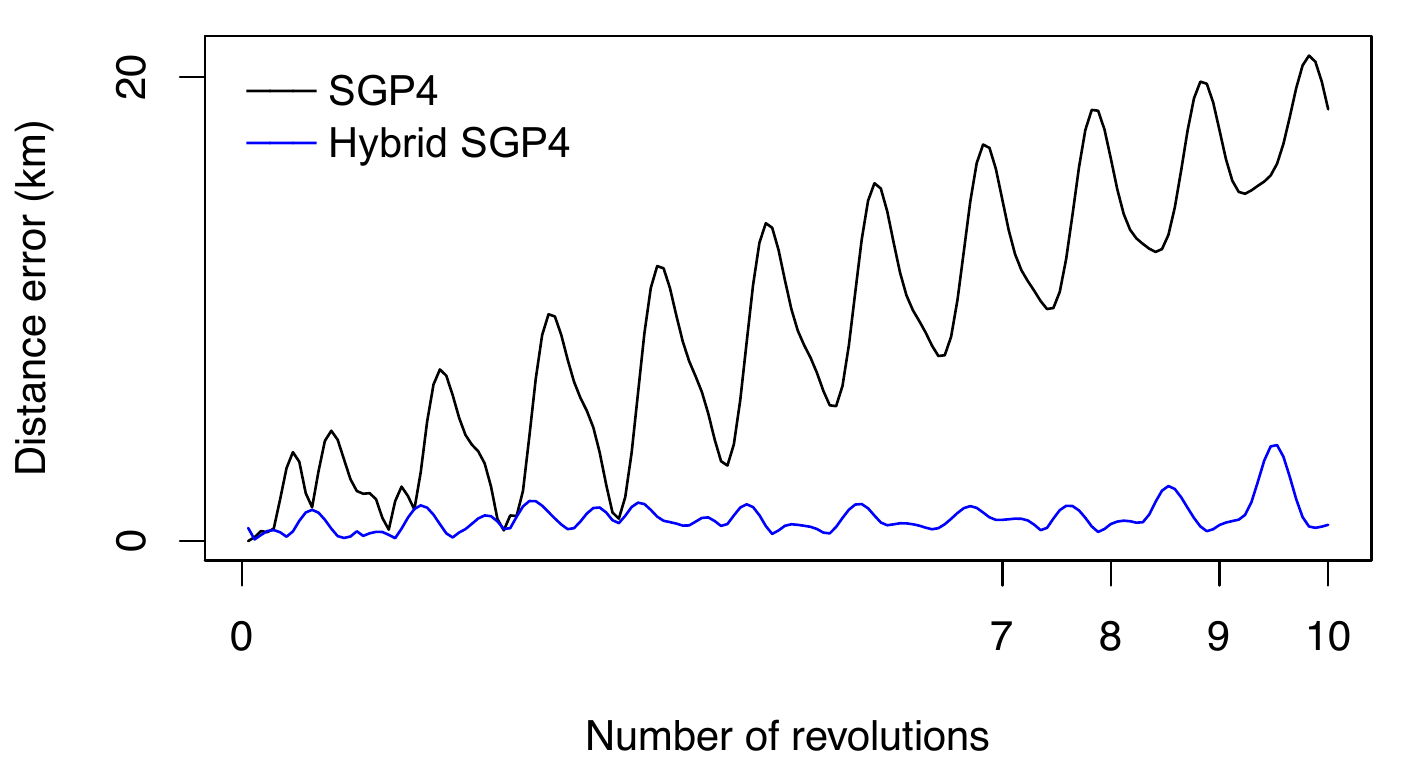}
\caption{Distance error of SGP4, in black, and the hybrid propagator, in blue, after a three-satellite-revolution propagation span}\label{fig4}
\end{figure}

A convenient way to evaluate how good a propagation is consists in translating the propagation errors into distance errors between the calculated position of the satellite and its real position, in this case determined through the AIDA numerical propagator. Fig. \ref{fig4} displays the distance error of SGP4 in black, and the hybrid SGP4 in blue. After the three revolutions during which we predict SGP4 errors, which correspond to approximately two propagation days, the maximum distance error for SGP4 is 20.92 km, whereas the hybrid propagator, in which the error correction has been applied, has a maximum distance error of only 5.98 km.

\begin{figure}[!htp]
\centering
\includegraphics[scale=.6]{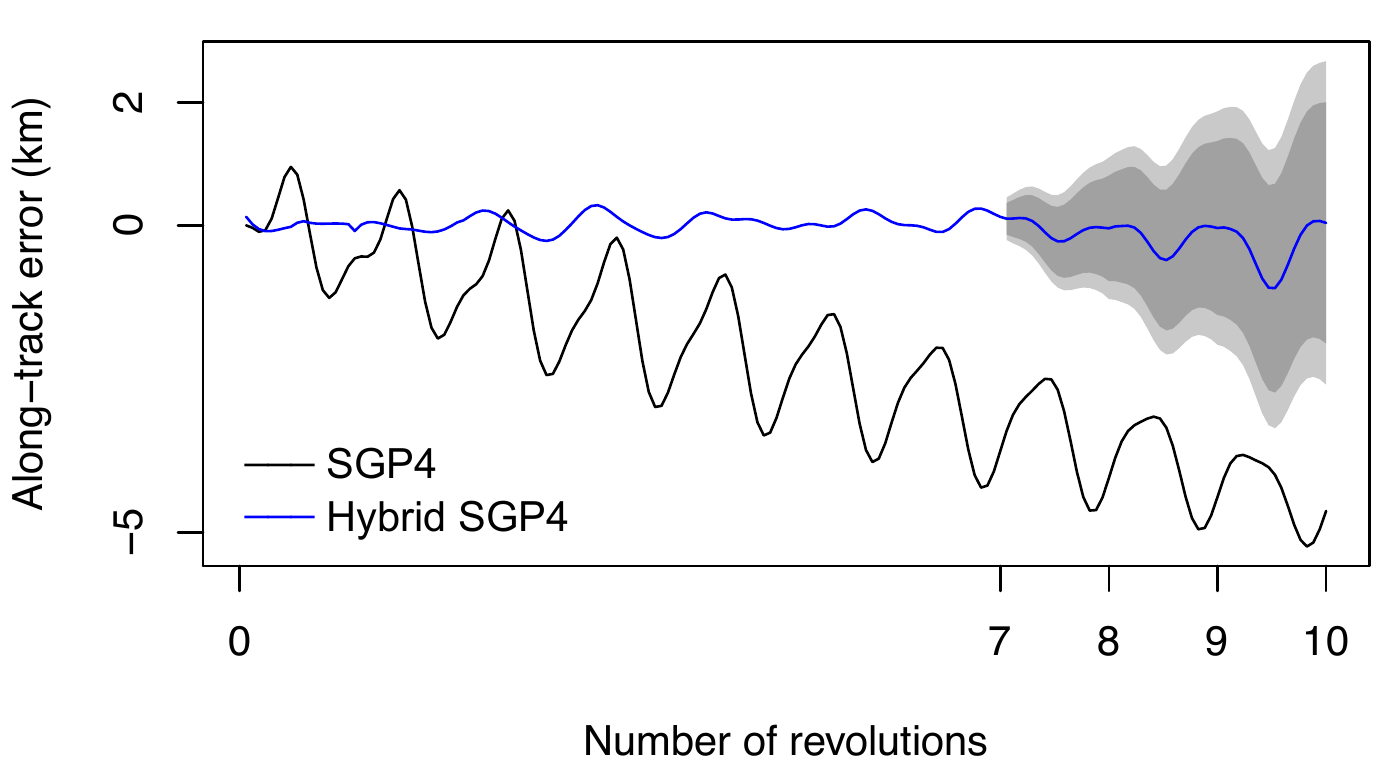}
\caption{Along-track error for SGP4, in black, and for the hybrid propagator, in blue. Shaded areas represent the 99\% and 95\% confidence intervals. Predictions start at revolution number 7.}\label{fig5}
\end{figure}

As can be expected, when we analyze the three orthogonal components of the position error, we verify that the main deviation takes place in the direction tangential to the orbit, namely the \textit{along-track error}. Fig. \ref{fig5} displays the along-track error for both SGP4, in black, and the hybrid SGP4, in blue, with respect to the real satellite position, accurately computed with the numerical propagator AIDA. This figure also shows the 99\% and 95\% confidence intervals, as dark- and light-gray halos, respectively. These confidence intervals have been calculated from the confidence intervals of the prediction of $\varepsilon^\theta$, shown in Fig. \ref{fig3}.

Finally, this study has been extended to the 53 different TLEs from the Galileo-8 satellite showed in Fig. \ref{fig2}. The same procedure has been followed in all the cases: the time series of the argument-of-the-latitude error $\varepsilon^\theta$ during the first seven satellite revolutions has been used for fitting the parameters of the model, and then, future errors have been predicted for the next three revolutions. Table \ref{tab1} presents some statistics for the distance errors for both SGP4 and the hybrid propagator HSGP4. As can be noticed, not only are HSGP4 errors smaller, but they also show a lower dispersion. The family of hybrid orbit propagators improves the accuracy of the classical SGP4, and is particularly good for short forecasting horizons.

\begin{table}
\caption{Statistics for the distance errors of SGP4 and the hybrid propagator HSGP4 (km)}\label{tab1}
\begin{center}
\begin{tabular}{l @{\quad\quad} c @{\quad\quad} c}
\hline\rule{0pt}{12pt}
& SGP4 error & HSGP4 error \\[2 pt]
\hline\rule{0pt}{12pt}
Minimum   & $5.98$  & $5.98$\\
1st quartile & $13.17 $ & $5.98$ \\
Median & $20.89$& $5.98 $  \\
Mean & $23.72 $& $6.95  $  \\
3rd quartile  & $31.45$ & $7.52 $\\
Maximum  & $51.65 $ & $12.06$\\[2pt]
\hline
\end{tabular}
\end{center}
\end{table}

\section{Conclusions}

The hybrid methodology for orbit propagation consists in complementing the approximate solution of a base propagator with a correction based on the time series of the propagator past errors, generated by means of statistical methods or machine learning techniques. It allows improving the accuracy of any base propagator, irrespective of its type, with a very light increment in the computational burden.

One of the most convenient statistical techniques for this purpose is the exponential smoothing method. We use it in order to create a model from the base-propagator past errors, and later to predict future errors.

In this study, we make use of the state-space formulation of the exponential smoothing method. Its main advantage lies in the fact that it allows applying the maximum likelihood method, which, by considering the error terms as Gaussian noise during the fitting process of the exponential-smoothing model parameters, allows determining the confidence interval of the predictions.

Knowing the confidence interval of the predictions allows propagating the uncertainties, which is necessary in order to determine the collision probabilities of space objects.

The study has been performed taking the well-known SGP-4 as the base propagator, and applying it to the propagation of Galileo-type orbits.

\subsubsection{Acknowledgments.} This work has been funded by the Spanish State Research Agency  and the European Regional Development Fund under Project ESP2016-76585-R (AEI/ERDF, EU).

\bibliographystyle{splncs04}
\bibliography{references}

\end{document}